\documentclass[10pt,aps,prl,twocolumn,notitlepage,superscriptaddress,preprintnumbers,longbibliography]{revtex4-1}

\usepackage{amsmath,amssymb,amsfonts}
\usepackage{bm}
\usepackage{graphicx,color}
\usepackage{verbatim}
\usepackage{float}
\usepackage{dcolumn}
\usepackage{natbib}
\usepackage{hyperref}
\usepackage{mdframed}

\newcommand{\avg}[1]{\langle #1\rangle}

\newcommand{\ea}{\emph{et al.} }
\newcommand{\ER}{Erd\"os-R\'enyi }
\newcommand{\BA}{Barab\'{a}si-Albert }

\begin{document}

\title{Fragility and anomalous susceptibility of weakly interacting networks}

\author{Giacomo Rapisardi}\email{giacomo.rapisardi@imtlucca.it}
\affiliation{IMT School for Advanced Studies, 55100 Lucca -- Italy}
\author{Alex Arenas}
\affiliation{Departament d'Enginyeria Inform\`{a}tica i Matem\`{a}tiques, Universitat Rovira i Virgili, 43007 Tarragona -- Spain}
\author{Guido Caldarelli}
\affiliation{IMT School for Advanced Studies, 55100 Lucca -- Italy}
\affiliation{Istituto dei Sistemi Complessi (CNR) UoS Sapienza, 00185 Rome -- Italy}
\affiliation{European Centre for Living Technology, Universit\`a di Venezia ``Ca' Foscari'', 30124 Venice -- Italy}
\author{Giulio Cimini}
\affiliation{IMT School for Advanced Studies, 55100 Lucca -- Italy}
\affiliation{Istituto dei Sistemi Complessi (CNR) UoS Sapienza, 00185 Rome -- Italy}

\begin{abstract}
Percolation is a fundamental concept that brought new understanding on the robustness properties of complex systems. 
Here we consider percolation on weakly interacting networks, that is, network layers coupled together by much less interlinks than the connections within each layer. 
For these kinds of structures, both continuous and abrupt phase transition are observed in the size of the giant component. 
The continuous (second-order) transition corresponds to the formation of a giant cluster inside one layer, and has a well defined percolation threshold. 
The abrupt transition instead corresponds to the merger of coexisting giant clusters among different layers, and is characterised by a remarkable uncertainty in the percolation threshold, 
which in turns causes an anomalous trend in the observed susceptibility. We develop a simple mathematical model able to describe this phenomenon 
and to estimate the critical threshold for which the abrupt transition is more likely to occur. 
Remarkably, finite-size scaling analysis in the abrupt region supports the hypothesis of a genuine first-order phase transition. 
\end{abstract}

\maketitle

Percolation theory is a very successful framework for understanding a broad range of critical phenomena taking place on networks, 
such as robustness to failures or attacks and spreading of diseases or information, and for unveiling the common principles underlying these processes \cite{stauffer2014introduction,RevModPhys.80.1275}. 
In this context, multilayer networks have been shown to exhibit critical percolation properties which are different from what is observed 
for a single isolated network---namely, a single continuous phase transition \cite{PhysRevLett.85.5468,PhysRevE.64.026118} 
whose properties depend on the kind of process \cite{radicchi2015break} and on the network features \cite{PhysRevE.66.036113,castellano2010threshold}. 
Indeed, the presence of interconnections between the network layers can give rise to supercritical phenomena such as abrupt or multiple phase transitions. 
Discontinuous percolation transitions have been extensively reported in the case of interdependent networks---that is, 
two (or more) networks whose nodes are interconnected by dependency links, 
such that the removal of a node in a network causes the instantaneous removal of the dependent nodes in the other networks 
(see for instance \cite{buldyrev2010catastrophic,son2012percolation,baxter2012avalanche}). 
Our focus here is instead on interacting networks (or network of networks), in which the connections between the network layers 
are ordinary links that thus take part in the percolation process. A system of this kind is therefore equivalent to a single modular network, characterised by a percolation threshold 
that is typically lower than in homogeneous networks---with a giant cluster appearing for a smaller total number of links \cite{leicht2009percolation}. 
A case of particular interest arises when the interaction between the network layers is weak, meaning that there is a sufficiently small number of interlinks between network layers, 
so that the removal of a few of them can easily separate the network layers into isolated modules \cite{shai2015critical}. 
This setup is common for neural systems, and therefore of major relevance to understand the resilience of neural processing \cite{Hoppensteadt:1997:WCN:261133}. 
Weakly interacting networks are characterised by a mixed percolation phase, in which only one or some of the network layers do percolate \cite{dickison2012epidemics,melnik2014dynamics}. 
In particular, Colomer-de-Sim\'on \& Bogu\~n\'a \cite{colomer2014double} identified multiple percolation transitions when the coupling between the different layers vanishes in the thermodynamic limit. 
In order to account for the emergence of coexisting percolating clusters, Faqeeh \ea \cite{faqeeh2016emergence} developed a modular message passing approach. 
In any event, the appearance of these coexisting clusters in weakly interacting networks is a fundamental source of error for percolation theory. 
In this work we develop a simple mathematical framework that allows estimating the most likely critical threshold at which the merging of coexisting clusters occurs in weakly interacting networks. 
Moreover, we characterise the percolation process in terms of a powder keg: due to the scarcity of the interlinks, the aggregation of the coexisting giant clusters is delayed, 
therefore giving rise to an abrupt percolation transition.

\section*{Results}

\subsection*{The intrinsic powder keg of weakly interacting networks}

\begin{figure*}
\centering
\includegraphics[width=1.50\columnwidth,angle=0]{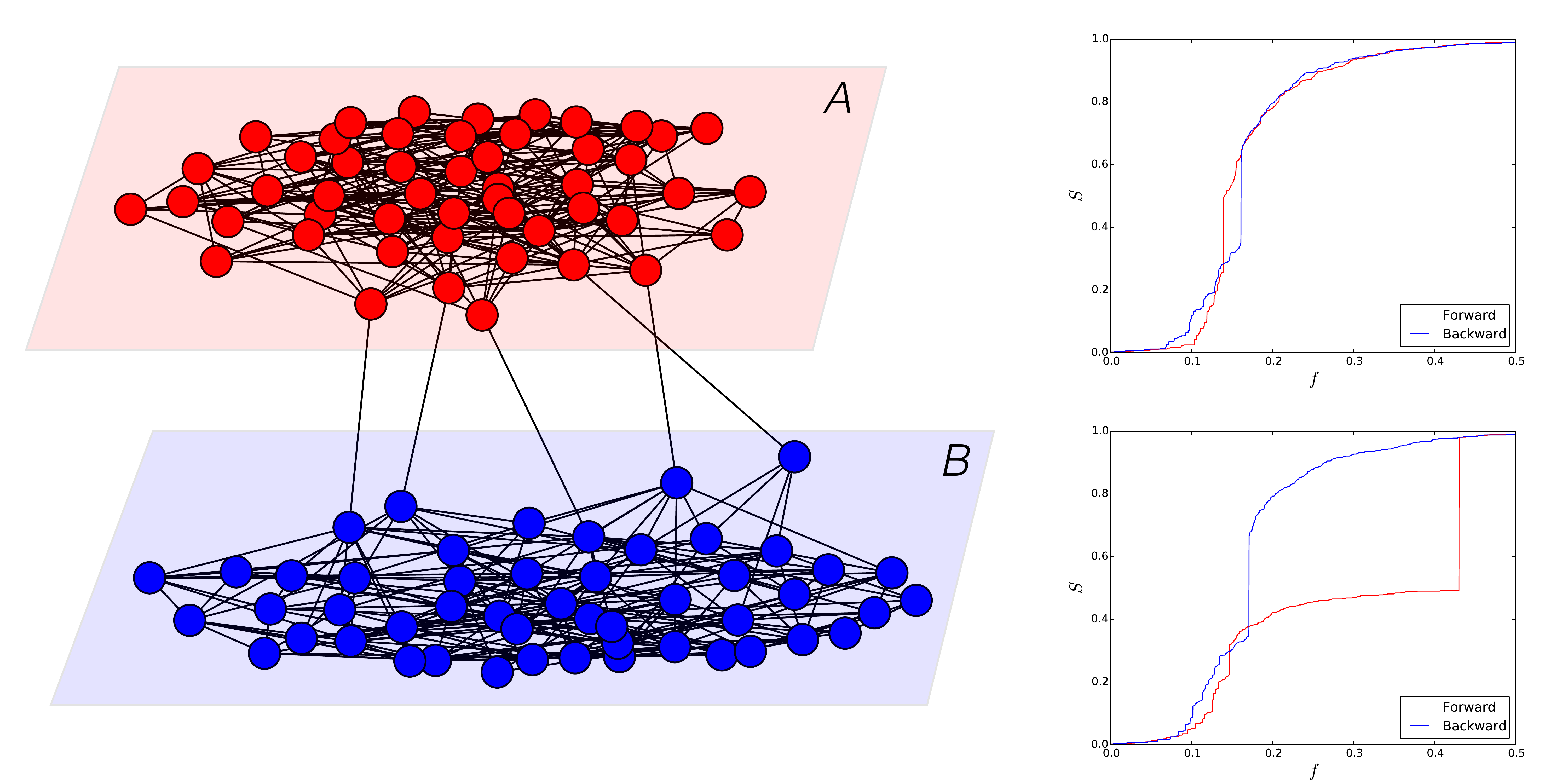}
\caption{{\bf Single realisation of bond percolation on a weakly interacting network.} 
Left panel: pictorial representation of two weakly interacting networks $A$ and $B$, in which the interconnection links $I$ are much less than the intra-layer links. 
Right panel: two different instances of the percolation process on an interacting network composed by two \ER layers ($N=500$ nodes and average degree $k=10$ each) 
connected by $I=5$ interconnection links. Each realisation is obtained as follows. Starting from an empty network, links are first randomly added ({\em forward}) up to half the total number of links, 
and then randomly removed ({\em backward}) until the network is empty again. The hysteresis cycle appearing in both cases are remarkably different, 
because of the large variability of the percolation threshold.}\label{fig:ist}
\end{figure*}

To illustrate the percolation properties of weakly interacting networks, we consider as in Figure \ref{fig:ist} two layers 
$A$ (with $N_A$ nodes and average degree $k_A$) and $B$ (with $N_B$ nodes and average degree $k_B$), that are interconnected by a small number $I$ of links ($I \ll \min\{N_Ak_A,N_Bk_B\}$). 
The bond percolation process consists in retaining each link of the system with occupation probability $f$ and otherwise removing it. 
To simulate the process, we use the method proposed by Newman and Ziff \cite{PhysRevLett.85.4104}: for each realisation, 
we start from a system configuration with no connections, and then sequentially add links in a random order. $f$ is thus the fraction of links added to the system. 
In such a situation, we may observe large jumps for the order parameter $S$, that is, the size of the giant cluster spanning both layers. 
These jumps can be understood as resulting from the addition of one of the $I$ interlinks {\em after} the formation of the two giant clusters $S_A$ and $S_B$ of layer $A$ and $B$, respectively. 
Indeed, differently from what happens for standard percolation, when such interlink is about to be added the two giant clusters already contain a number of nodes that is proportional to the system size. 
According to the definition of Friedman \ea \cite{Friedman2009powderkeg}, this configuration corresponds to a {\em powder keg}, which is ``ignited'' as soon as that interconnection is added 
causing a discontinuous percolation transition. Note that if a system is initialised as a powder keg, then even a random link addition rule causes a discontinuous transition: 
as in our case, the formation of the giant cluster spanning both layers is not hindered by specific link selection rules \cite{Achlioptas2009,grassberg2011explosive,Nagler2011,Nagler2012,DSouza2015}, 
but is naturally delayed by the structure of the interconnections itself. However, the absence of any particular link selection criteria causes a large uncertainty for the percolation threshold.

\begin{figure*}
\centering
\includegraphics[width=1.75\columnwidth,angle=0]{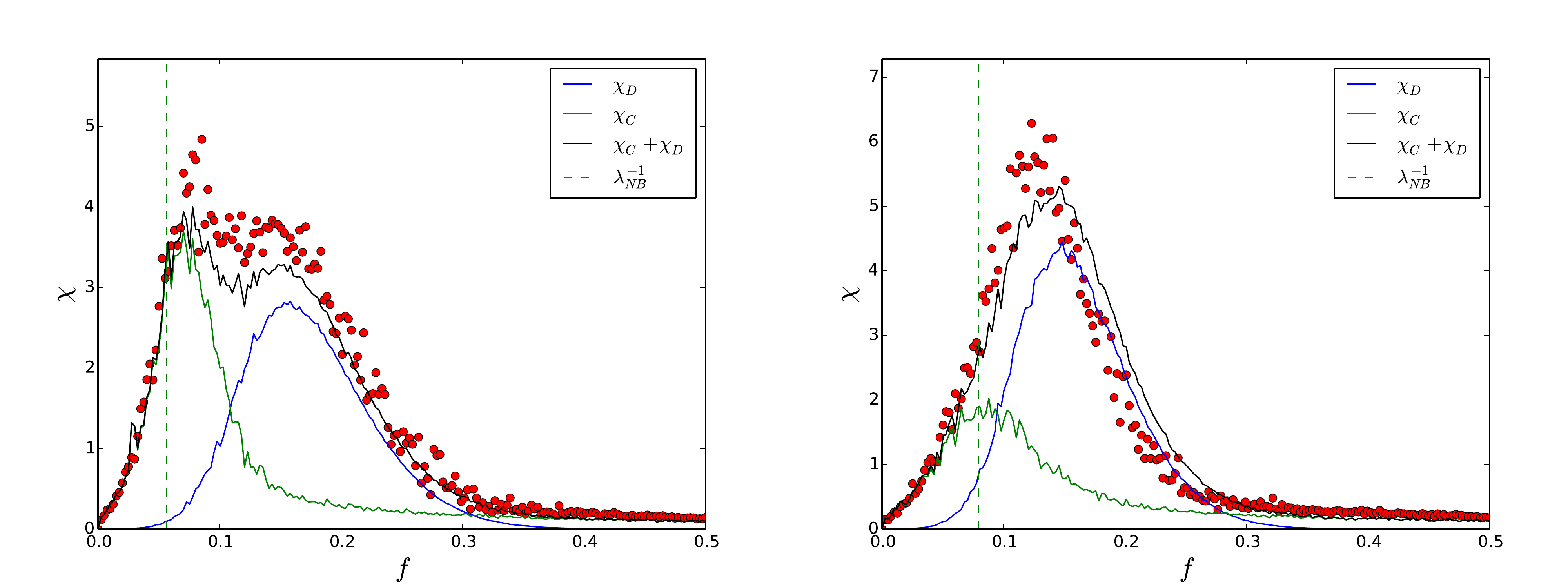}
\caption{{\bf Susceptibility of air transportation networks.} 
We consider the duplex formed by the transportation network Lufthansa-Ryainair (left panel) and Lufthansa-Easyjet (right panel) \cite{Cardillo2013}, 
in which each layer is made up of airports (nodes) and flight routes (links) operated by a company. 
The layers are characterised by $N_{LH}=106$ and $\avg{k}_{LH}=4.604$, $N_{FR}=128$ and $\avg{k}_{FR}=9.391$, $N_{U2}=99$ and $\avg{k}_{U2}= 6.202$. 
The interconnection links in each case are the airports in which both companies operate: we have $I_{LH-FR}=36$ and $I_{LH-U2}=51$. 
Red dots denote numerically computed values of $\chi$ from $400$ realisations of the bond percolation process, 
$\chi_D$ is given by eq. \eqref{eq:chi} and $\chi_C$ is the susceptibility of the corresponding non-interacting system. 
$\lambda_{NB}$ is the leading eigenvalue of the non-backtracking matrix of the network, whose inverse is a good approximation 
for the percolation threshold of sparse networks \cite{Karrer2014perc}.}\label{fig:eudata}
\end{figure*}

\subsection*{Model for the anomalous susceptibility and percolation threshold}

In order to gain a more quantitative insight on the described phenomenology, we start by defining the probability $P_I$ 
that at least one of the $I$ interconnections is added and actually connects the two giant clusters $S_A$ and $S_B$ \cite{faqeeh2016emergence}: 
\begin{equation}\label{eq:bridge}
P_I = 1 - \left[ 1- \left( \frac{N}{N_A}S_A \right) \left( \frac{N}{N_B}S_B \right)f \right]^{I} =
1 - \left[ 1- \frac{N}{\mu}S_AS_Bf \right] ^{I},
\end{equation}
where $N=N_A+N_B$ is the total number of nodes, the normalisation coefficients before $S_A$ and $S_B$ respectively denote their maximum size $N_A/N$ and $N_B/N$, 
and $\mu = \frac{N_AN_B}{N_A+N_B}$ is the {\em reduced} number of nodes (equivalently to the concept of reduced mass for the classical two-body problem). 
Without loss of generality, we set the percolating thresholds $f_A$ and $f_B$ of the individual layers $A$ and $B$ respectively such that $f_A<f_B$ 
(the degenerate case $f_A=f_B$ is reported below and discussed in the Methods section). This implies that on average and for layers of the same nature 
we have $S_A>S_B$ for any given value of $f$ such that both clusters exist. 
Hence, for $f>f_B$, the percolation cluster $S$ of the whole system is either that of layer $A$ if $S_A$ and $S_B$ are not connected, 
or abruptly jumps to $S_A+S_B$ provided that $S_A$ and $S_B$ are connected---which happens with probability $P_I$. In formulas,
\begin{equation}\label{eq:S}
S = \left\{\begin{array}{ll}
  S_A+S_B & \text{ with probability } P_I\\
  S_A & \text{ otherwise}
\end{array}\right.
\end{equation}
Overall, we have a first continuous transition at $f_1 = f_A$ (the standard percolation transition when layer $A$ percolates), and a second discontinuous transition at $f_2$ when
layer $B$ percolates and at least one active interconnection is established between the two layers. Yet, because of the dichotomy characterising the outcome of the process for $f\simeq f_2$, 
the average value $\avg{S}=S_A+S_BP_I$ is not representative at all of what happens in the system. We thus study the behaviour of the susceptibility $\chi=N\mbox{Var}(S)/\avg{S}$ \cite{colomer2014double}. 
For $f>f_B$ each layer has its own percolating cluster, and thus the only contribution to $\chi$ comes from the Bernoulli trial described by eq. \eqref{eq:S}: 
\begin{equation}\label{eq:chi}
\chi_D=N\frac{S_B^2P_I(1-P_I)}{S_A+S_BP_I}.
\end{equation}
Note that $\chi_D$ gives a non-vanishing contribution to the total susceptibility $\chi$ only in the weakly interacting regime, that is, when the Bernoulli trial of eq. \eqref{eq:S} is not trivial. 
Indeed, eq. \eqref{eq:bridge} tells us that for $I\to0$ (as for the case of disconnected layers) we have $P_I\to0$, and when $I$ is very large (as is the case of strongly connected layers, see the Methods section) 
we have $P_I\to1$. In both cases $\chi_D\to0$. For fixed $I$, however, $\chi_D$ achieves its maximum for the value of $f$ which maximises the uncertainty of the Bernoulli trial, 
at which the discontinuous jump of $S$ is more likely to occur. We thus identify $f_2$ with the value for $f$ which maximises $\chi_D$.

\subsection*{Real and artificial networks}

\begin{figure*}
\centering
\includegraphics[width=2.00\columnwidth,angle=0]{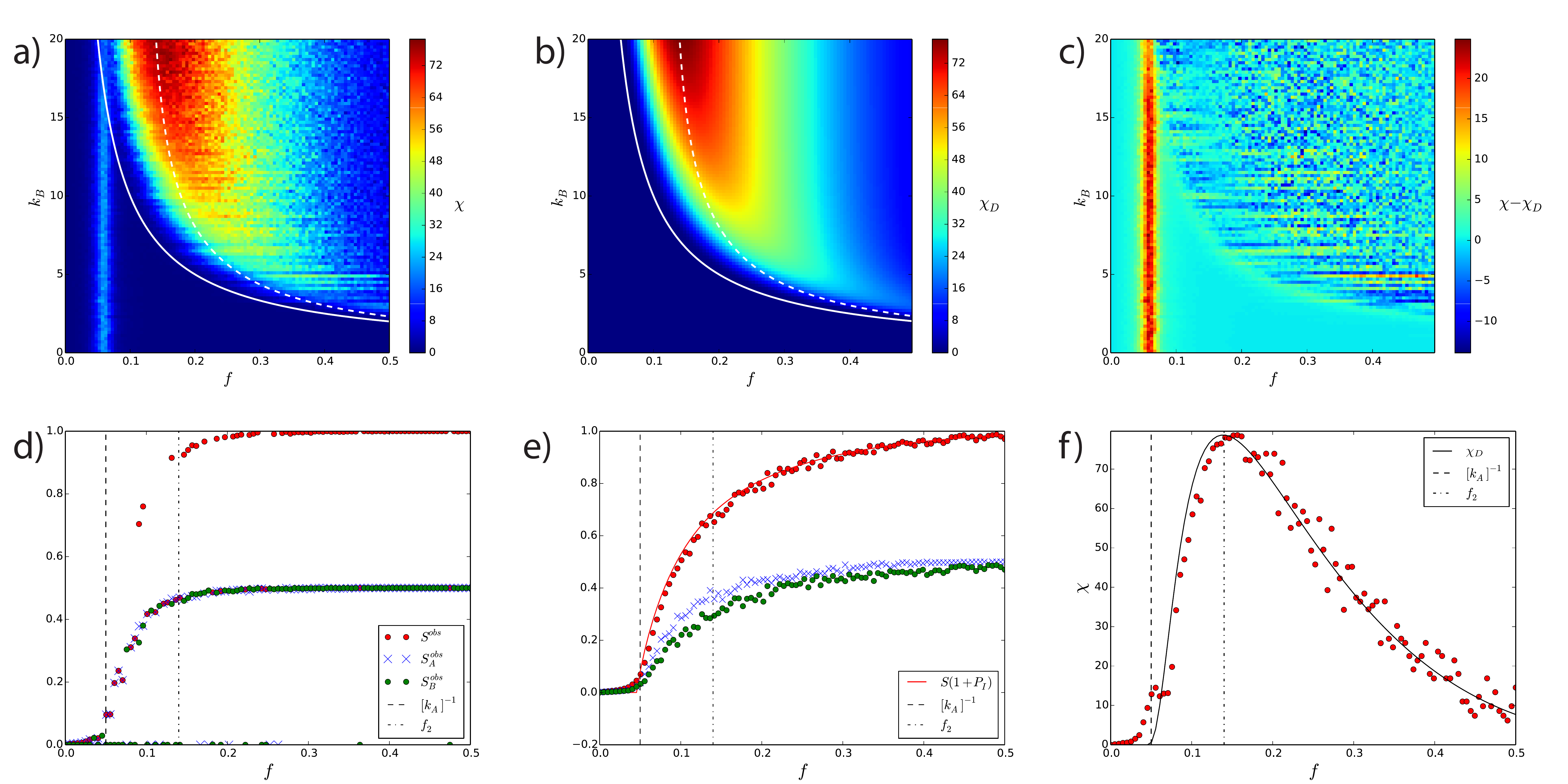}
\caption{{\bf Susceptibility of weakly interacting \ER networks.} 
Panels (a-c) are heat maps of the susceptibility of two \ER coupled networks with $N_A=N_B=500$ nodes, $k_A=20$, $I=5$ and $k_B$ varying from $0$ to $k_A$. 
Panel (a) reports $\chi$ obtained from numerical simulations of the percolation process, whereas, panel (b) reports $\chi_D$ from numerical solutions of eq. \eqref{eq:chi}. 
In both cases, the continuous white line gives the relation $f=k_B^{-1}$ that marks the boundary for the region in which percolating clusters exist in both layers, 
and the dashed white line reports values of $f_2$ as given by eq. \eqref{eq:f2}. Panel (c) reports the difference between the two values, which is high only in the vertical strip corresponding to $\chi_C$. 
Panels (d-f) instead report $S$ and $\chi$ for the degenerate case of two \ER interacting layers with $N_A=N_B=500$ and $k_A=k_B=20$, and $I=5$. 
In all three cases, the dashed vertical line denotes the percolation threshold of the individual layers, whereas, the dashed-dotted line marks $f_2$ as derived from eq. \eqref{eq:f_2simm}.
Panel (d) reports a single realisation of the process and Panel (e) reports averages of the same process over $300$ realisations. 
In both cases, red dots are the observed values of $S$,  blue crosses and green dots are the observed values of $S_A$ and $S_B$ respectively, 
and the red line gives the numerical estimate of $\avg{S}$ derived from eq. \eqref{eq:S}. Panel (f) finally reports the observed susceptibility (red dots) 
averaged over $300$ realisations of the process, as well as the numerical value of $\chi_D$ (continuous black line).}\label{fig:heats}
\end{figure*}

These simple mathematical arguments are indeed able to capture the behaviour of the susceptibility both in real and model networks. 
We first consider in Figure (\ref{fig:eudata}) the duplex (two-layer multiplex) formed by a pair of coupled air transportation networks, 
where each layer consists of the airports (nodes) and flight routes (links) operated by a given company, and the interlinks are the airports served by both companies. 
We see that the susceptibility of the two individual layers $\chi_C$ cannot capture the observed behaviour of $\chi$ computed numerically. 
The difference between $\chi$ and $\chi_C$ is instead very well represented by $\chi_D$. 

A more precise assessment of our methodology is given by considering two \ER weakly interacting networks 
with the same number of nodes $N_A=N_B$ and average degrees $k_A$ and $k_B$. 
In this case it is possible to derive an analytic approximation for $f_2$, since $S_A$ and $S_B$ have a known analytic form in the thermodynamic limit. 
We get (see the Methods section for details):
\begin{equation}\label{eq:f2}
f_2=\frac{1-2^{-1/I}}{1-\exp[-k_B(1-2^{-1/I})]}. 
\end{equation}
The specific case $k_A=k_B=k$ leads to a more accurate transcendental equation (see again Methods):
\begin{equation}\label{eq:f_2simm}
f_2 = \frac{ 1- (2 - \sqrt{2})^{1/I}}{ \left\{ 1 - \exp \left[ -k\sqrt{f_2(1- (2 - \sqrt{2})^{1/I})} \right] \right\}^2 },
\end{equation}
that can be easily solved numerically.
As shown in Figure \ref{fig:heats}, in the case of two weakly interacting \ER networks with different average degrees, 
the numerical evaluation of $\chi_D$ by means of eq. \eqref{eq:chi} fits very well the observed anomalous susceptibility, 
and the numerical solution of eq. \eqref{eq:f2} gives with good approximation the position of the maximum of $\chi$. 
In the degenerate case $k_A=k_B$, eq. \eqref{eq:f_2simm} provides an even better approximation for the maximum of $\chi$. 
Analysing single realisation of the percolation process, we confirm that $f_2$ marks the region in which $S$ is subject to discontinuous jumps. 
However, this discontinuous behaviour is lost by averaging the outcomes of the percolation process over many realisations, 
for which $S$ becomes $\avg{S}=S_A + P_IS_B$ which fails to represent the outcome of the process.

\subsection*{Finite size scaling}

\begin{figure*}
\centering
\includegraphics[width=1.50\columnwidth,angle=0]{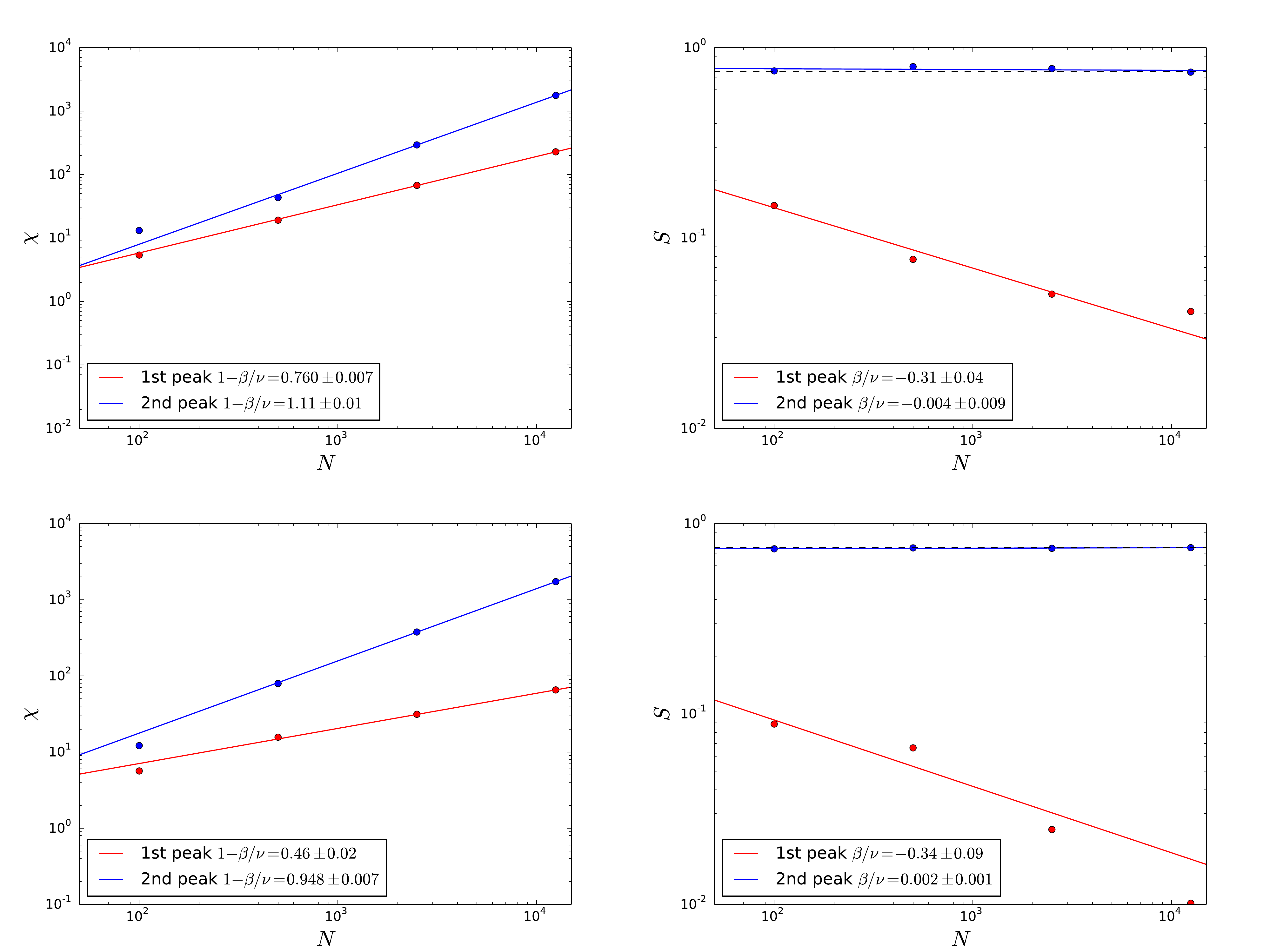}
\caption{{\bf Finite size scaling analysis.} 
Top panels report the case of two weakly interacting \ER layers with $k_A = 20$, $k_B = 10$ and $I = 5$, for different size $N$.
From (b) we see that while $S(f_1)$ shows a power law decay with exponent $\beta / \nu = 0.31\pm0.04$ (which is consistent with the mean-field values $\beta = 1$ and $\nu=3$), 
$S(f_2)$ does not scale with $N$. Accordingly to those values, from (a) we can verify the different divergence rates for the two peaks of the susceptibility, 
and in particular we see that the divergence of $\chi(f_2)$ is almost linear. 
Bottom panels instead report the case of two weakly interacting \BA layers with $m_A = 20$, $m_B = 10$ and $I = 5$, for different size $N$.
Again we see that while $S(f_1)$ and $\chi(f_1)$ show a scaling behaviour ruled by the topology of the layers, 
$S(f_2)$ and $\chi(f_2)$ show the same behaviour of the \ER case: the one characteristic of first-order phase transitions.}\label{fig:finite_size}
\end{figure*}

We conclude our study with a finite-size scaling analysis carried out for the cases of two coupled \ER layers and two coupled \BA layers with different average connectivities. 
For each of the two settings we considered networks made of two layers of size $N_{A,B}$ equal to 100, 500, 2500, 12500. 
According to standard percolation theory, the maximum of the susceptibility diverges around the critical value $f_c$ according to the power law 
$\chi(f_c) \sim N^{1 - \beta / \nu}$, while for the relative size of the giant component we have $S(f_c)\sim N^{-\beta / \nu}$ \cite{colomer2014double,radicchi2015break}.
Our analysis, reported in Figure \ref{fig:finite_size}, shows that the scaling properties around $f_1$ and $f_2$ are significantly different. 
While $S(f_1)$ and $\chi(f_1)$ exhibit the usual power law scaling typical of second-order phase transitions (with different exponents according to the two different topologies of the network layers), 
$S(f_2)$ does not scale with $N$ in both examples, which implies $\beta / \nu = 0$. This in turn implies $\chi(f_2)\sim N$. 
These particular scaling properties, that is the failure of finite size scaling relations and the extensive character of the susceptibility, 
are a clear trademark of a first order phase transition \cite{Radicchi2009explosive}.

\section*{Conclusion}

To sum up, in this work we have studied the bond percolation properties of weakly interacting networks. 
This class of systems encompass the important cases of multilayer/modular networks with very sparse connections within the layers/modules. 
We reported the existence of discontinuous jumps in the relative size of the giant component $S$, 
happening since the percolating cluster of the sparser layer can give either a full or zero contribution to the giant cluster of the whole system. 
Furthermore we observed that in this case the abrupt transition does not have a definite threshold, 
but can occur for a wide range of values of the bond occupation probability. This causes an anomalous behaviour of the susceptibility, which we captured using simple probabilistic arguments. 
We successfully tested our predictions in both synthetic and real systems. 
Finally, from finite-size scaling analysis we showed that the critical behaviour of both $S$ and $\chi$ in the abrupt region 
exhibits the features of a genuine first-order phase transition.
Our work can have important applications in characterising the fragility of weakly interacting structures such as multiplex transportation networks, 
as well as in describing epidemic processes on networks with metapopulation structures \cite{WANG20122689,6580178,PhysRevE.88.012809,PhysRevE.90.032806,doi:10.1063/1.4990592}.

\medskip

A. A. acknowledges the Spanish MINECO (grant no. FIS2015-71582-C2-1). A. A acknowledges funding also from ICREA Academia and the James S. McDonnell Foundation. 
G.C. and G.C. acknowledge support from the EU H2020 projects DOLFINS (grant no. 640772) and CoeGSS (grant no. 676547).

\newpage

\section*{Methods}

In order to derive the analytic approximations presented in both eqs. \eqref{eq:f2} and \eqref{eq:f_2simm} 
in the case of two \ER layers of the same size ($N_A=N_B=N/2$), we start from the implicit form of $S_A$ and $S_B$ in the thermodynamic limit:
\begin{equation}\label{eq:S_A_B}
S_X = \tfrac{1}{2}\left(1 - e^{-2fk_XS_X}\right)
\end{equation}
with $X=\{A,B\}$. The above expression is obtained from the usual equation for a single \ER network, namely $S=1-e^{-fkS}$, using the substitution $S\to 2S$ 
(as $S_X$ refers to only one layer with half of the $N$ nodes). We thus obtain the same solution of the single network scaled by a factor $1/2$, as well as the same percolation threshold $f_X=k_X^{-1}$. 
The value of $f$ which maximises $\chi_D$ of eq. \eqref{eq:chi} for fixed $S_A$ and $S_B$ is given by the following implicit equation
\begin{equation}\label{eq:f2_chi}
[1-4f_2S_AS_B]^{I}=1+\frac{S_A - \sqrt{S_A^2+S_AS_B}}{S_B},
\end{equation}
where both $S_A$ and $S_B$ are functions of $f_2$ according to eq. \eqref{eq:S_A_B}.
Not that eq. \eqref{eq:f2_chi} returns $f_2=1-(2-\sqrt{2})^{1/I}$ in the limit $S_{A,B}\to1/2$. 
This regime corresponds to the case $I\ll k_{A,B}$, for which we can safely assume
that both layers will fully percolate before the activation of at least one interconnection link as $f$ increases,
leading to a value of $f_2$ which does not depend on $k_A$ nor $k_B$.
Since eq. \eqref{eq:f2_chi} is difficult to handle, we can approximate $f_2$ with the values that maximises $\mbox{Var}(S)$ instead of $\chi_D$. 
For a Bernoulli trial we simply have $P_I(f_2)=1/2$, implying $f_2=[1-2^{-1/I}]/[4S_AS_B]$. 
With the further assumption $S_A \simeq 1/2$ (hence when layer $A$ has already percolated) we have $f_2=[1-2^{-1/I}]/[2S_B]$. 
Using eq. \eqref{eq:S_A_B} we finally get the analytic solution presented in eq. \eqref{eq:f2}:
$$f_2=\frac{1-2^{-1/I}}{1-\exp[-k_B(1-2^{-1/I})]}. $$
In the limit $S_B\to1/2$ this expression simplifies to $f_2=1-2^{-1/I}$, which is very close to the value that maximises $\chi_D$ in the same regime.

In the degenerate case $k_A=k_B=k$, we have $S_A = S_B = S_X$ which leads to the simpler expressions:
\begin{equation}
\avg{S} = S_X(1 + P_I), \qquad \chi_D = N \frac{S_XP_I(1-P_I)}{1+P_I},
\end{equation}
and thus the value of $f$ which maximises $\chi_D$ at fixed $S_X$ is simply given by the implicit expression $P_I(f_2) = \sqrt{2} -1$, 
implying $f_2=[1-(2 - \sqrt{2})^{1/I}]/[4S^2]$. Plugging the latter in eq. \eqref{eq:S_A_B} yields eq. \eqref{eq:f_2simm}
$$f_2 = \frac{ 1- (2 - \sqrt{2})^{1/I}}{ \left\{ 1 - \exp \left[ -k\sqrt{f_2(1- (2 - \sqrt{2})^{1/I})} \right] \right\}^2}.$$

\begin{figure}
\centering
\includegraphics[width=1.00\columnwidth,angle=0]{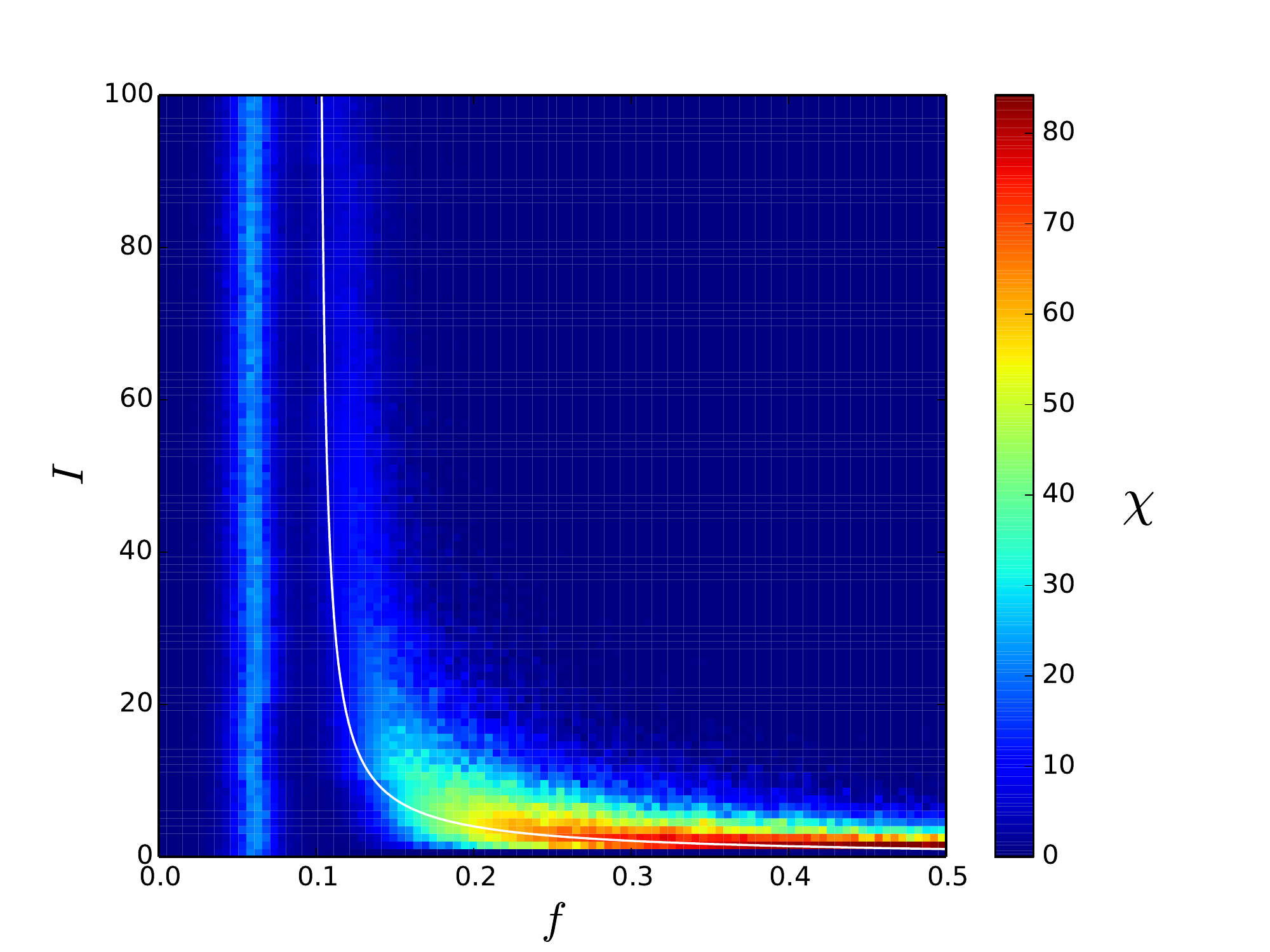}
\caption{{\bf The case of strongly connected layers.} Heatmap of the susceptibility $\chi$ for two \ER layers of $N=500$ with $k_A = 20$, $k_B=10$ 
and $0<I<100$. For every fixed value of $I$, $\chi$ is averaged over $400$ realisations of the bond percolation process. 
The continuous white line represents the theoretical prediction from eq. \eqref{eq:f2}, 
which for large values of $I$ converges to $k_B^{-1}=0.1$.}\label{fig:heatQ}
\end{figure}

We finally consider the case of strongly interacting \ER layers, that we define by $I \geq \max\{k_A,k_B\}$. 
As shown in Figure \ref{fig:heatQ} (where $k_A>k_B$), as soon as $I > k_A$ the height of the second peak 
drastically decreases, while the corresponding value of $f_2$ approaches $k_B^{-1}$, that is, the percolation threshold of the weak layer. 
The fact that $f_2\to k_B^{-1}$ is obtained by taking the limit $I^{-1}\to0$ in eq. \eqref{eq:f2}. 
Indeed, in this regime $P_I\simeq 1$ as soon as the percolating cluster appears in layer $B$: 
the process bears no uncertainty related to the interconnections, therefore contribution of $\chi_D$ vanishes 
and $\chi$ simply becomes that of the ordinary percolation process for the \ER layer $B$.

\end{document}